\documentclass[12pt]{article}
\usepackage{pic04}
\usepackage{hyperref}
\usepackage{url}
\usepackage{graphicx}
\usepackage{bm}

\begin{document}

\newcommand{\cmpi}{\mbox{\large$\bm{\langle}$}\ensuremath{\sin(\phi+\phi_S)}
        \mbox{\large$\bm{\rangle}$}\ensuremath{_{UT}^\pi}}
\newcommand{\smpi}{\mbox{\large$\bm{\langle}$}\ensuremath{\sin(\phi-\phi_S)}
        \mbox{\large$\bm{\rangle}$}\ensuremath{_{UT}^\pi}}
\newcommand{\cmh}{\mbox{\large$\bm{\langle}$}\ensuremath{\sin(\phi+\phi_S)}
        \mbox{\large$\bm{\rangle}$}\ensuremath{_{UT}^h}}
\newcommand{\smh}{\mbox{\large$\bm{\langle}$}\ensuremath{\sin(\phi-\phi_S)}
        \mbox{\large$\bm{\rangle}$}\ensuremath{_{UT}^h}}
\newcommand{\mb}{\mathbf}
\newcommand{\st}{T}
\newcommand{\xbj}{x}
\newcommand{\de}{{\rm\,d}}

\title{\bf TRANSVERSITY AND SPIN STRUCTURE FUNCTIONS}
\author{H. E. Jackson \\
{\em Argonne National Laboratory, Argonne, Illinois 60439}}
\maketitle

%
%
%
%
%
%
\vspace{4.5cm}
%

\baselineskip=14.5pt
\begin{abstract}
Measurements of single-spin asymmetries for semi-inclusive electro-production
of pions and kaons in deep-inelastic scattering 
with transverse target polarization open a new window
on the transverse quark and gluon structure of the nucleon. The first 
experimental results from such measurements as well as experiments in
progress are discussed. Properties of the the spin-dependent Collins
fragmentation function and prospects for extracting the transversity 
are reviewed and evidence for nonzero Sivers asymmetries as manifestations
of quark orbital angular momentum is evaluated.
\end{abstract}

\newpage
\baselineskip=17pt
\section{Introduction}
A complete description of the quark structure of the proton at 
leading order in deep-inelastic scattering (DIS) requires three
flavor-dependent parton distribution functions (PDF's). The most
familiar of these is the unpolarized distribution function 
$q(x,Q^{2})$. It describes the quark momentum distribution
at infinite momentum. 
Here $x$ is the dimensionless Bjorken scaling variable which
is the momentum fraction carried by a parton 
and $-Q^2$ is the four momentum transfer. The
first moment of $q(x,Q^2)$ provides a measure of the quark vector charge,
{\it i.e.} $\left\langle PS|\overline{\psi}\gamma^{\mu}\psi|
PS\right\rangle = \int_{0}^{1} \, dx(q(x)-\overline{q}
(x))$. It has been throughly studied \cite{pumplin,martin} 
and is well known for all 
flavors. In the quark parton model, the $F_2$ structure function is
given by $F_{2}(x)=x\sum_{q}e^{2}_{q}q(x)$. The second PDF is the
longitudinal polarized distribution function $\Delta q(x,Q^{2})$ 
which describes the longitudinal helicity distribution of the quarks
in a proton polarized parallel to its momentum. Its first moment
determines the axial charge of the quarks, {\it i.e.} $\left\langle PS
|\overline{\psi}\gamma^{\mu}\gamma_{5}\psi|
PS\right\rangle = \int_{0}^{1} \, dx(\Delta q(x)+\Delta\overline{q}
(x))$. In terms of the polarized PDF's, the polarized proton structure 
function is given by $g_{1}(x)=0.5\sum_{q}e^{2}_{q}\Delta (x,Q^{2})$.
In recent years much has much has been learned about $g_1$ from studies
of polarized DIS \cite{filippone}. The third PDF is 
the transversity distribution function
$\delta q(x,Q^{2})$ which measures the quark helicity distribution in 
a proton polarized perpendicular to the proton momemtum at infinite 
momentum. It is related to the quark tensor charge through its first 
moment, {\it i.e.} $\left\langle PS
|\overline{\psi}\sigma^{\mu\nu}\gamma_{5}\psi|
PS\right\rangle = \int_{0}^{1} \, dx(\delta {q}_{1} (x)+
\delta {\overline{q}}_{1}(x))$. Until now its properties have been unobserved.

For non-relativistic quarks, $\Delta q$ and $\delta {q}$ would be identical
since by means of commuting rotations and Euclidean boosts one can 
convert a longitudinally polarized proton into a transversely polarized 
proton at infinite momentum. However, because the internal motion of the
quarks is relativistic, this is not true and a comparison of the two PDFs 
will reflect the relativistic character of quark motion in the proton. In 
contrast to its chiral-even partners, $q$ and $\Delta q$, the transversity
distribution is chiral-odd. This property of transversity makes its 
measurement difficult since hard QCD and electroweak processes preserve
chirality. It decouples from inclusive DIS and other deep inelastic processes.
However, it is of considerable interest because of its unique properties.
The $Q^2$ evolution of $\delta {q}(x,Q^{2})$ is much simpler than that
of its leading order partners, because it does not 
couple to gluons. It is a valence
quantity. Lattice gauge calculations can provide reliable estimates of its
first moment, the quark tensor charge.   

Chirality can be conserved in DIS processes involving transversity by
coupling it to a second chiral-odd function, in the case of Drell-Yan
to the transversity of the beam particle, or in the case of semi-inclusive
hadron production to a fragmentation function which also has a chiral-odd 
structure. Consequently, reactions involving at least two hadrons are
required to explore the properties of transversity. 
Indeed, transversity was first introduced by 
Ralston and Soper \cite{ralston} in their treatment of the 
production of muon pairs in polarized Drell-Yan production.
\begin{figure}[t]
\centerline{\includegraphics[width=13cm]{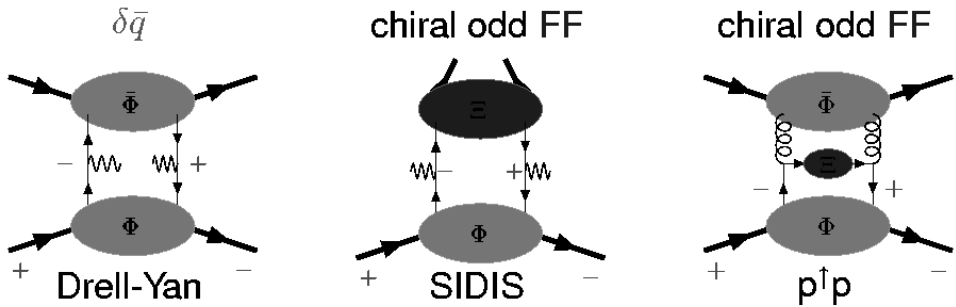}}
 \caption{\it
      Deep-inelastic scattering processes involving a chiral-odd process
that couples to transversity to conserve chirality.
    \label{exfig1} }
\end{figure}
Perhaps the most striking example of such reactions came almost 10 years
ago from the Fermilab experiment E704 which reported measuring a large
analyzing power $A_N$ in the inclusive production of pions from a 
transversely polarized proton beam of 200 GeV incident on an unpolarized
target \cite{bravar}. Very strong azimuthal asymmetries were reported for
positive and neutral pions which were of opposite sign when measured for
negative pions.

Two explanations were developed for these results. The observable
$A_N$ is odd under the application of naive time reversal and was assumed
therefore to arise from the non-perturbative part of the reaction amplitude.
Assuming factorization of the amplitude as shown in Figure \ref{exfig1}, 
either a T-odd fragmentation function or a T-odd distribution function must 
be involved. In one case the measured analyzing power was postulated to be
proportional to the product of transversity $\delta q$ times a Collins
fragmentation function $H_{1}^{\perp}$ which is T- and chiral-odd. In the
second case the T-odd chiral-even distribution function $f_{1T}^{\perp}
(x,k_{T})$ first postulated by Sivers \cite{sivers}
is coupled to the usual unpolarized fragmentation function $D_{q}^{h}$. Either
a pure Collins amplitude \cite{ansel1} or a pure Sivers amplitude
\cite{ansel2} provides a complete fit to the data. 

\section{Azimuthal spin asymmetries}

\begin{figure}[t]
\begin{center}
\includegraphics[height=8cm,angle=-90]{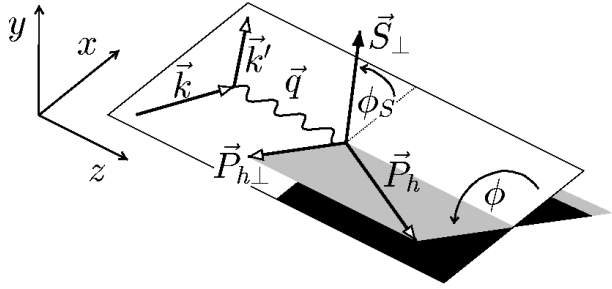}
\end{center}
\caption{\label{kinematics} The definitions of the azimuthal
angles of the hadron production plane and the axis of the 
relevant component $\vec{S}_\perp$ of the target spin,
relative to the plane containing the momentum $\vec{k}$ ($\vec{k}'$) of the
incident (scattered) lepton.
Explicitly,
$\phi = \frac{\vec{q}\times\vec{k}\cdot\vec{P}_h}
             {|\vec{q}\times\vec{k}\cdot\vec{P}_h|}
\cos^{-1}{\frac{\vec{q}\times\vec{k}\cdot\vec{q}\times\vec{P}_h}
               {|\vec{q}\times\vec{k}||\vec{q}\times\vec{P}_h|}}$
and $\phi_S = \frac{\vec{q}\times\vec{k}\cdot\vec{S}_\perp}
                   {|\vec{q}\times\vec{k}\cdot\vec{S}_\perp|}
\cos^{-1} {\frac{\vec{q}\times\vec{k}\cdot\vec{q}\times\vec{S}_\perp}
                {|\vec{q}\times\vec{k}||\vec{q}\times\vec{S}_\perp|}}$,
where $0<\cos^{-1}<\pi$.
 }
\end{figure}

The Collins mechanism \cite{collins} 
is of special interest because it provides a direct
probe of $\delta q$. It produces a correlation  
in the fragmentation process between the axis
of transverse target spin and the vector ${\vec{P}_{\pi}}\times \vec{q}$ where
${\vec{P}_{\pi}}$ is the momentum of final state hadron and $\vec{q}$ is the 
virtual photon momentum. (See Figure \ref{kinematics}.) Because it is a
chiral-odd correlation it combines with $\delta q$ to provide an observable
which is accessible in semi-inclusive DIS. The Collins fragmentation
function $H_{1}^{\perp}$ which describes this spin-momentum correlation is
chiral-odd and odd under naive time reversal (T-odd), {\it i.e.} time 
reversal without interchange of the initial and final states. This ``Collins
function'' arises from the interference of amplitudes with different
imaginary parts and gives rise to hadron single-spin asymmetries. It
describes the influence of the transverse polarization on the
momentum component $\vec{P}_{h\perp}$ of the hadron transverse to the
virtual photon direction which will effect its distribution in the azimuthal
angle $\phi$ shown in Figure \ref{kinematics}. 
The characteristic signature for the Collins effect in
semi-inclusive scattering of an unpolarized lepton beam by a transversely 
polarized target producing a pseudoscalar meson is a $sin(\phi+\phi_{S})$
variation in the azimuthal distribution \cite{boer}.

Azimuthal spin asymmetries can also be generated by the T-odd Sivers
distribution function $f^{\perp}_{1T}(x,k_{T})$ discussed in the
previous section, which describes a correlation between the 
transverse polarization of the target nucleon and the $\vec{p}_T$
of the struck quark. This $\vec{p}_T$ survives fragmentation to be inherited
by the hadron $\vec{P}_{h\perp}$. The familiar unpolarized fragmentation
function $D_1$ describes the fragmentation process.  Because this
mechanism is related to a forward scattering amplitude in which the 
helicity of the target nucleon is flipped ($N^{\Rightarrow}q\;\rightarrow
\;N^{\Leftarrow}q$), the angular momentum of the unpolarized quark plays
a role. The polarization state of the virtual photon and the orientation of
the lepton scattering plane are irrelevant to the Sivers effect. The 
characteristic signature for this mechanism is a target spin asymmetry
with a $sin(\phi-\phi _{S})$ variation.

Measurements of spin asymmetries with transverse target polarization, permit
one to use the variation in the two azimuthial angles $\phi$ and $\phi_S$ 
to distinguish the Collins and Sivers effects. The quantity measured is
the two dimensional asymmetry
\begin{eqnarray} \label{eq:Aphi}
{A}^h_{UT}(\phi,\phi_S)&=& \frac{1}{|S_T|} 
\frac{\left(N_h^\mathbf{\uparrow}(\phi,\phi_S) 
  - N_h^\mathbf{\downarrow}(\phi,\phi_S)\right)}
{\left(N_h^\mathbf{\uparrow}(\phi,\phi_S) 
  + N_h^\mathbf{\downarrow}(\phi,\phi_S)\right)}\, \\
  &=&{A}^h_{C}sin(\phi +\phi_{S}) + A^{h}_{S}sin(\phi-\phi_{S}) \nonumber
\end{eqnarray}
where $N_h^{\mathbf{\uparrow}(\mathbf{\downarrow})}(\phi,\phi_S)$
is the semi-inclusive luminosity-normalized yield in that target
spin state, and $\phi_S$ always indicates the spin direction
of the $\mathbf{\uparrow}$ state.  Measurement with an unpolarized
beam and transversely polarized target is indicated by the
UT subscripts. This asymmetry is then fit with a
sum of contributions from two sinusoidal dependences as show above. 
Monte Carlo simulations confirm that extraction of both contributions 
is made from this fit without measurable cross-contamination even 
in the case that their magnitudes in the acceptance are very different.

\section{Experimental measurements}

\subsection{HERMES}

\begin{figure}[t]
\begin{center}
\includegraphics[width=14cm]{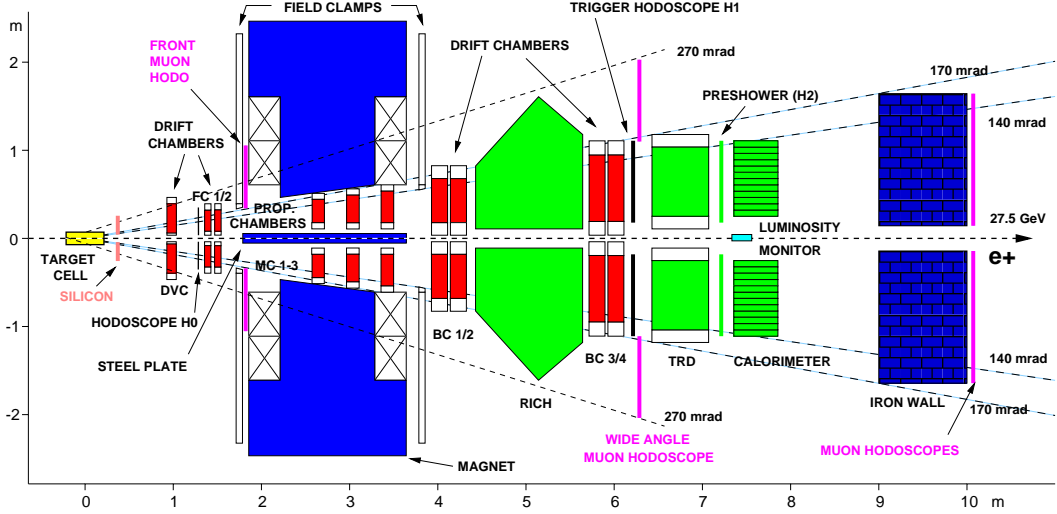}
\end{center}
\caption{\label{spectrom} The HERMES spectrometer.
 }
\end{figure}
The studies of $\delta q$ and $f_{1T}^{\perp}$ in the HERMES experiment
employ a transversely nuclear-polarized hydrogen gas target 
\cite{stock} which intercepts
the $E=27.6\;GeV$ beam of the HERA positron storage ring. The beam was 
unpolarized for these measurements. An open-ended cell is fed by an 
atomic-beam source based on Stern-Gerlach separation with hyperfine
transitions. The nuclear polarization of the atoms is flipped at 60 s 
time intervals, and the average proton polarization $S_T$ was 
0.78$\pm$0.04. Scattered beam leptons and coincident hadrons are 
detected by the HERMES spectrometer \cite{acker}. Its geometrical acceptance
covers the range 40 $<$ $|\theta _y|$ $<$ 140 mrad and $|\theta _x|$ $<$ 170
mrad where $\theta _x$ and $\theta _y$ are projections of the polar 
scattering angle. The scattered leptons are identified with 98\% 
efficiency and less than 1\% hadron contamination by means of an 
electromagnetic calorimeter, a transition-radiation detector, a preshower
scintillator counter, and a \v{C}erenkov detector. Charged pions are
identified by means of a dual-radiator ring-imaging \v{C}erenkov
detector \cite{akopov}.  

\begin{figure}[ht]
\includegraphics[width=8cm]{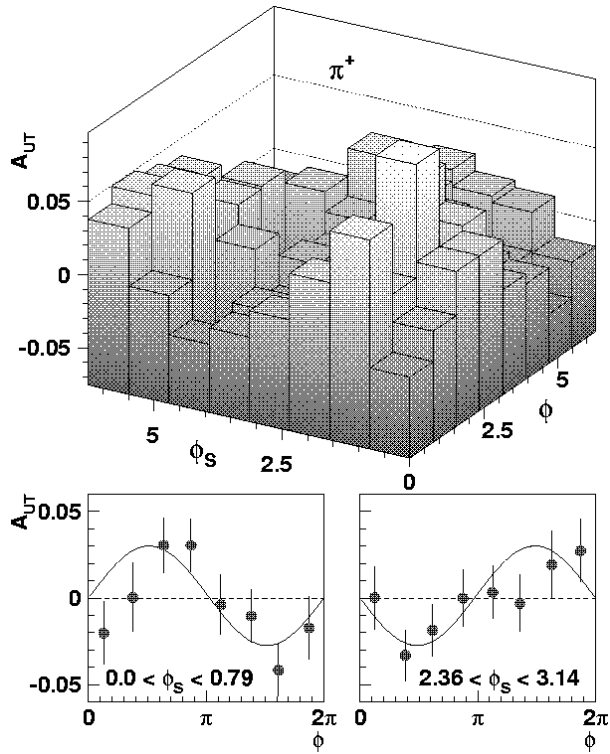}
\hspace{.5cm}
\begin{minipage}{5cm}\vspace{-10cm}
 \caption{\it
      Measured $\pi^+$ cross section asymmetries in transverse 
polarization averaged over the experimental acceptance as a function
of azimuthal angles shown in Figure \ref{kinematics}. Projections of
the upper panel are shown in the lower panels, selected from the 
indicated ranges in the other angle \label{exfig} }
\end{minipage}
\end{figure}

Events are accepted within the kinematic limits 
$W^{2}\,>$ 10 GeV$^2$, 0.1 $<\, y \, <
0.85$ and $Q^{2}\,>$ 1 GeV$^2$, where  $W$ is the invariant mass of the
photon-nucleon system and $y=(P\cdot q)/P\cdot k)$. Coincident hadrons
were required to be in the range 0.2 $< z <$ 0.7 and 
$\theta_{\gamma^{*}}\, >$ 0.02 rad, where $z=(P\cdot P_{h})/(P\cdot q)$
and $\theta_{\gamma^{*}}$ is the angle between the direction of the virtual
photon and that of the hadron. All hadrons detected for each event were 
included. Effects of acceptance, instrumental smearing, and QED radiation
were all found to be negligible in Monte Carlo simulations. Data for the 
single spin asymmetry $A^{\pi^{+}}_{UT}(\phi ,\phi_{S})$ presented in 
Figure \ref{exfig} show the expected sinusoidal two-dimensional behaviour.

\begin{figure}[hb]
\includegraphics[width=85mm]{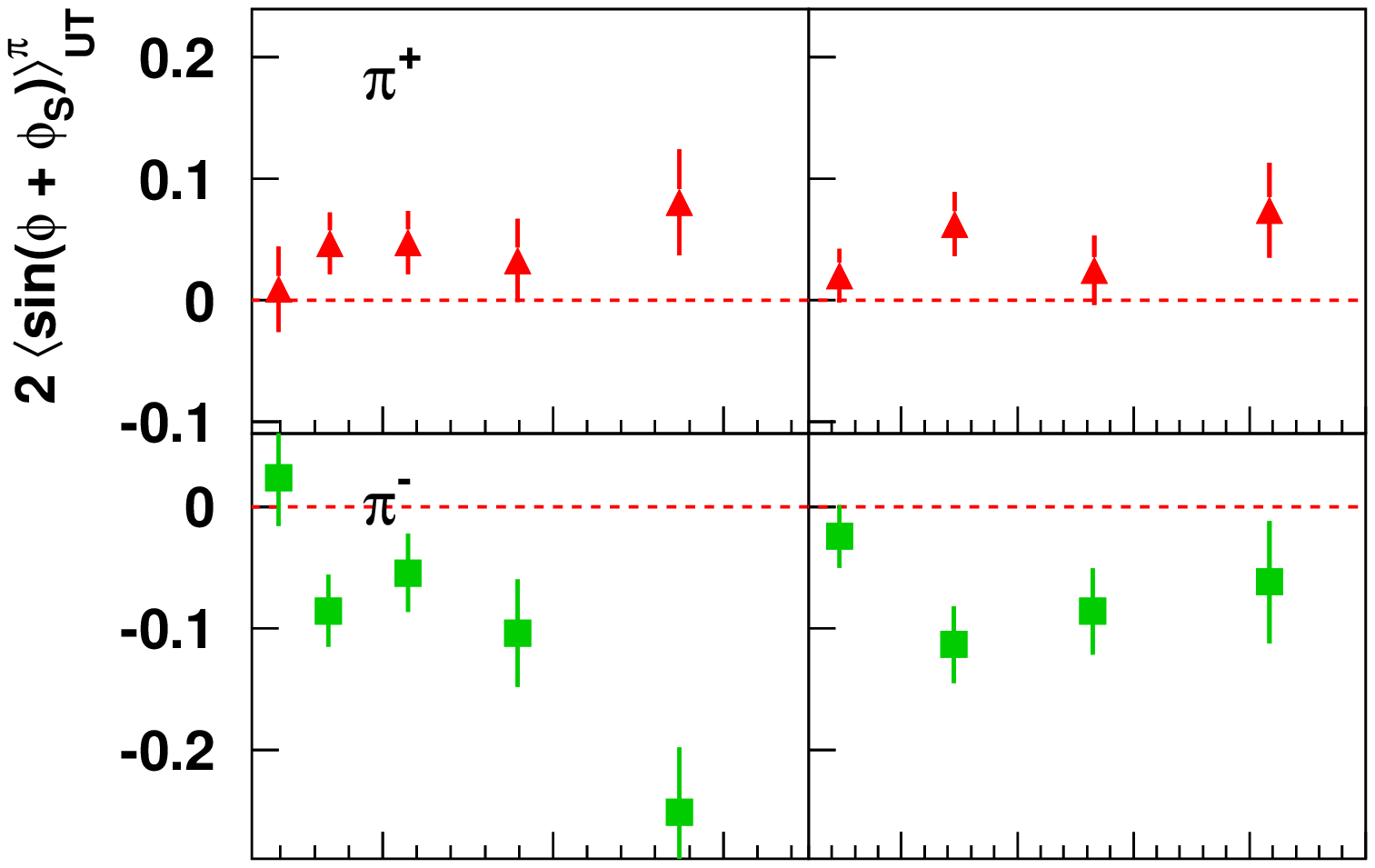} \hfill \\
\includegraphics[width=85mm]{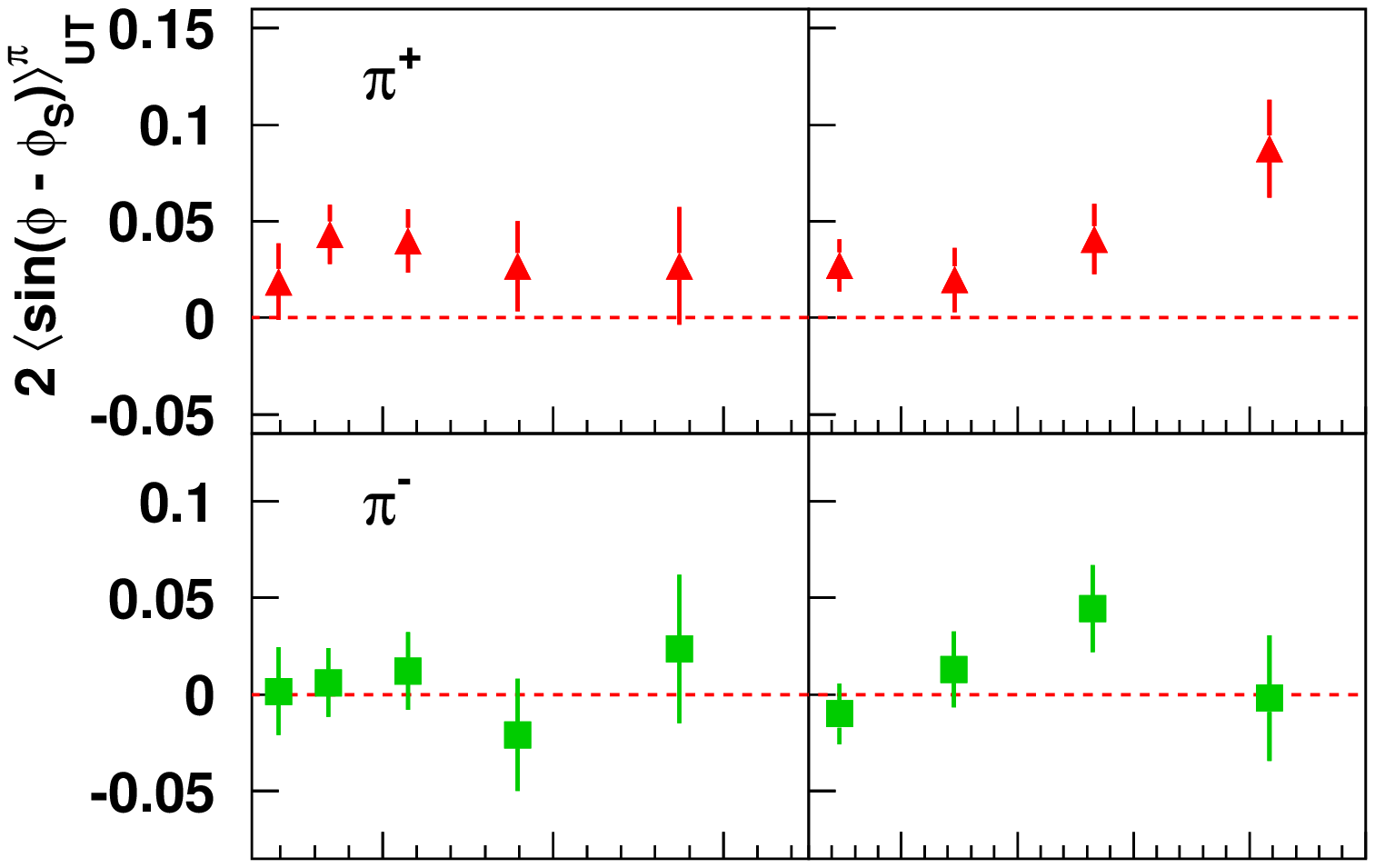} \hfill \\
\hspace*{2.5mm}\includegraphics[width=82.5mm]{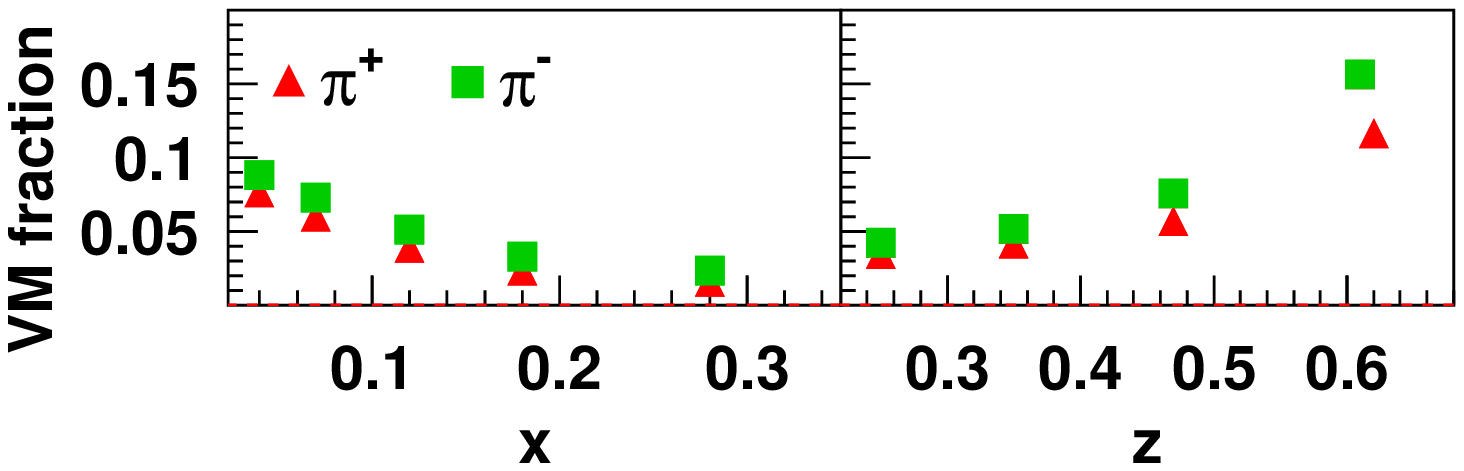} 
\hspace{0.9cm}
\begin{minipage}{5cm}\vspace{-14cm}
\caption{\label{fig:asymm} Virtual-photon Collins (Sivers) moments
for charged pions as labelled in the upper (middle) panel,
as a function of $x$ and $z$.
The error bars represent the statistical uncertainties.
In addition, there is a common 8\% scale uncertainty in the moments.
The lower panel shows the relative contributions to the data from
simulated exclusive vector meson production.
 }
\end{minipage}
\end{figure}

The HERMES analysis \cite{hermes} is based on the 
extraction of a Collins azimuthal
moment $\cmh$ and Sivers moment $\smh$ of the virtual-photon asymmetry
from the fit to a modified version of equation \ref{eq:Aphi} 
\begin{eqnarray} \nonumber
\frac{{A}^h_{UT}(\phi,\phi_S)}{2} &=& \cmh
\frac{B(\langle y\rangle)}
     {A(\langle x\rangle,\langle y\rangle)}
\sin(\phi+\phi_S) \\
&+& \smh \sin(\phi-\phi_S)\,.
\end{eqnarray}
where $B(y) \equiv (1-y)$,
$A(x,y) \equiv \frac{y^2}{2}+(1-y)(1+R(x,y))/(1+\gamma(x,y)^2)$, 
$R(x,y)$ is the ratio of longitudinal to transverse DIS cross sections,
$\gamma(x,y)^2 \equiv 2 M_p x/(E y)$, and $E$ is the lepton energy. 
The extracted asymmetries in the form
of azimuthal moments averaged over the experimental acceptance and selected
ranges of $x$ and $z$ of 0.023 $<$ $x$ $<$ 0.4 and 0.2 $<$ $z$ $<$ 0.7 are
shown in Figure \ref{fig:asymm}.
The corresponding mean values of the kinematic parameters are
$\langle x \rangle = 0.09, \langle y \rangle = 0.54,
\langle Q^2 \rangle = 2.41$\,GeV$^2,
\langle z \rangle = 0.36 $ and $ \langle P_{\pi\perp} \rangle = 0.41$\,GeV.
The bottom section of Figure \ref{fig:asymm} presents simulations based on
{\sc pythia6} \cite{PYTHIA6},
tuned for HERMES kinematics, of the fractions of the semi-inclusive
pion yield from exclusive production of vector mesons, the asymmetries
of which are poorly determined.

The results for the Collins moments show an unexpected behaviour. The 
averaged Collins moment for $\pi^+$
is positive at $0.021\pm0.007$(stat), while it is negative
at $-0.038\pm0.008$(stat) for $\pi^-$. The suggestion that transversity
distributions should resemble helicities distributions, at least in their
general trends is not born out by the data. To the extend that 
$\delta u$ is positive and $\delta d$ is negative the data is similar
to model predictions \cite{barone}.  However, the magnitude of
the negative $\pi^-$ moment appears to be at least as large as that 
for $\pi^+$. This trend becomes more pronounced as the magnitude of 
the transverse moments increase at large $x$ where valence quark effects
dominate, as they do in previously measured longitudinal spin asymmetries.
Unlike the case of $\pi^+$ where u-quark dominance is expected, large 
negative $\pi^-$ moments are a surprise, because neither quark flavor
dominates $\pi^-$ production, and one expects 
$|\delta d|<|\delta u|$ in analogy with $|\Delta d|<|\Delta u|$.
In addition, the Collins moments shown in the right portion of 
Figure \ref{fig:asymm} do not show the increase with increasing $z$
which has been predicted \cite{bacc} on the basis of a corresponding
$z$ dependence found for the Collins fragmentation function.

The HERMES results for the Sivers moment are very suggestive. The value 
found is positive and nonzero at $0.017\pm0.004$(stat) for $\pi^+$, while
the $\pi^-$ moment is consistent with zero: $0.002\pm0.005$(stat).
The large moment for $\pi^+$ appears to provide the first evidence in
leptoproduction for the T-odd Sivers parton distribution. Because
the $\pi^+$ moment is dominated by $up$ quarks, with the sign convention
which has been adopted in relating the azimuthal asymmetries to the
the parton distributions \cite{mulders} a positive asymmetry implies
a negative value for the Sivers function of this flavor. However, a 
substantial contamination of pions from $\rho ^0$ decay are present.
Studies to date of a small sample of exclusive $\rho ^0$ events in which
both pions are detected suggest that this asymmetry extracted for the
$\pi^+$ in the same manner as in the HERMES semi-inclusive analysis
also has a significant positive $\smh$ asymmetry which could complicate 
the interpretation of the data for the Sivers effect. 

\subsection{COMPASS}

\begin{figure}[t]
\begin{center}
\includegraphics[width=15cm]{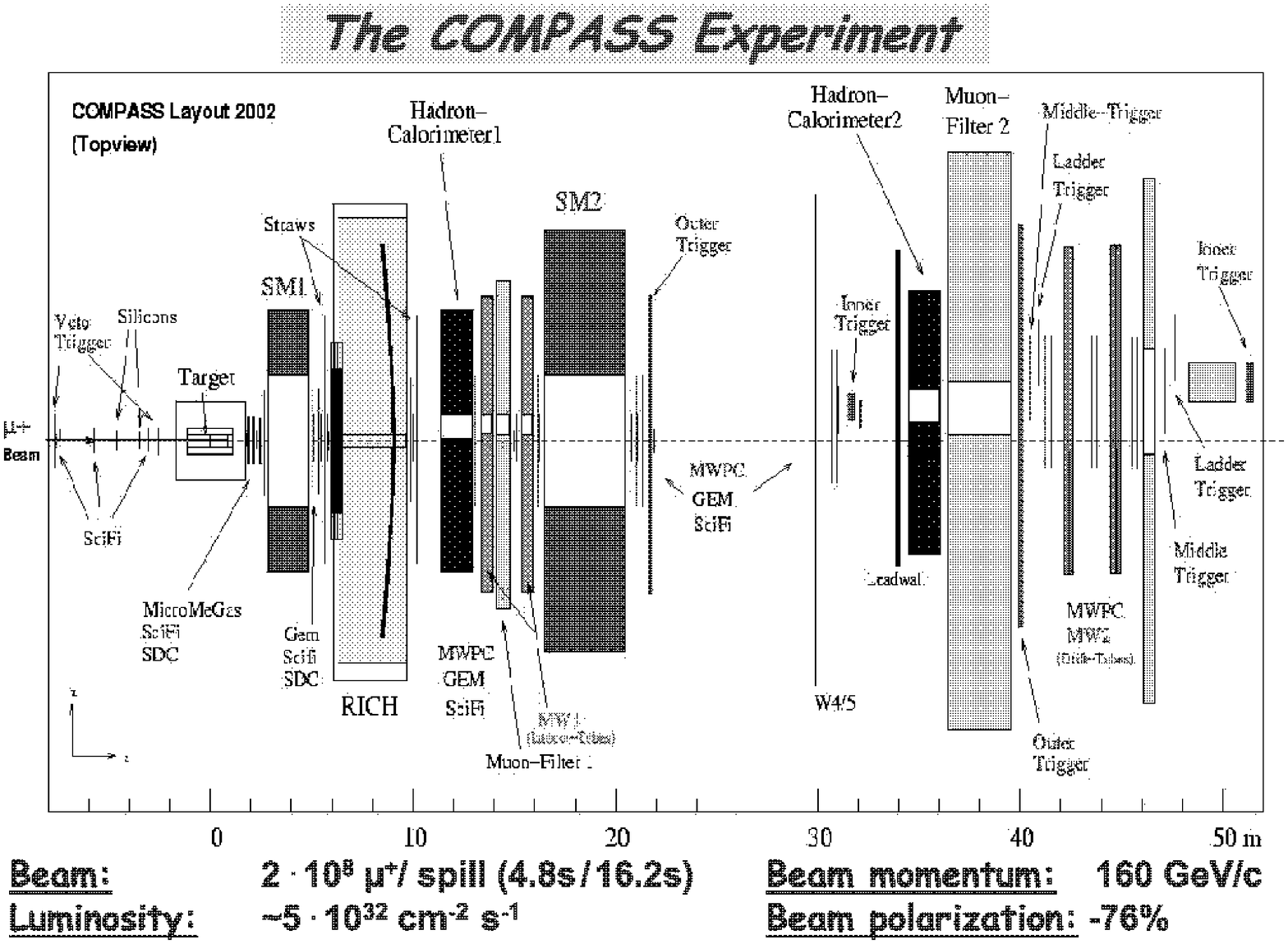}
\end{center}
\caption{\label{spectrom2} Top view of the COMPASS spectrometer.
 }
\end{figure}
The measurement of transversity is a major component of the physics
program of the COMPASS experiment which uses the CERN SPS muon beam.
The experiment has been run at a muon energy of 160 GeV. The beam 
intensity is 2$\cdot$10$^8$ muons per spill which is 4.5 seconds long.
A polarized $\vec{Li}\vec{D}$ target containing two simultaneously oppositely
transversely polarized sections is viewed by a two stage fixed target
magnetic spectrometer shown in Figure \ref{spectrom2}. To date, data 
reported results from events in which leading hadrons are accepted without 
further particle identification.

\begin{figure}[t]
\begin{center}
\includegraphics[width=15cm]{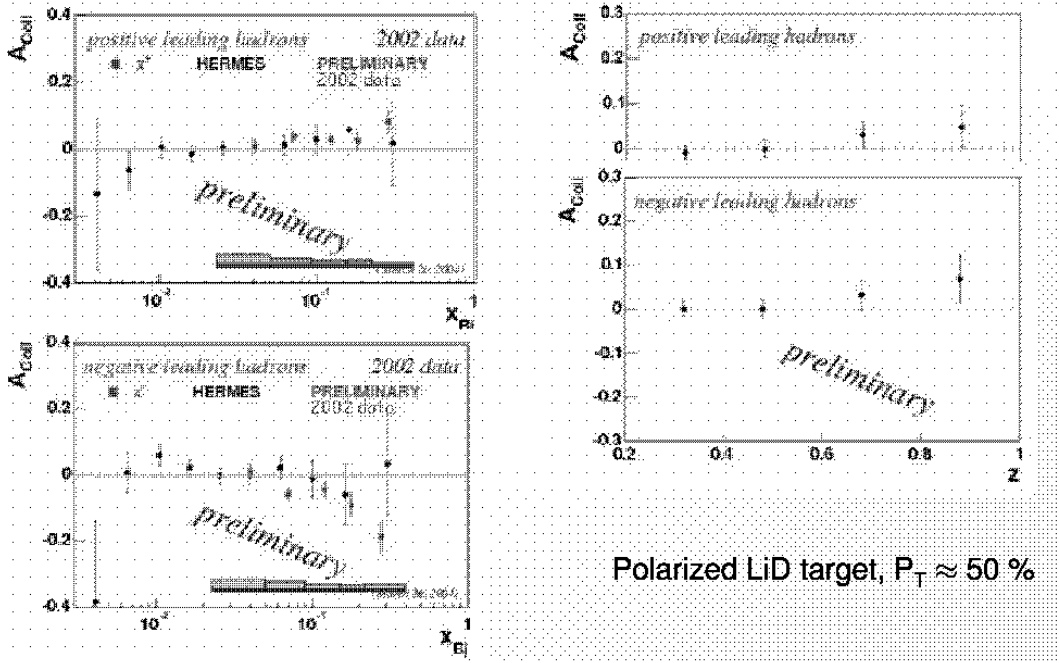}
\end{center}
\caption{\label{compass2} The Collins asymmetry as measured by COMPASS.
HERMES data are also shown for comparison.
 }
\end{figure}
Preliminary results for the Collins asymmetry as measured by COMPASS 
\cite{fischer} is
presented in Figure \ref{compass2}. The kinematic restrictions on the 
scattered muon give $Q^{2}\, >\, 1\, GeV^2$, 0.1 $<$ y $<$ 0.9, and 
w $>$ 5 GeV. Leading hadrons are accepted if $p_{T}>0.1 Gev/c$ and 
$z>0.25$. While the statistical precision is still less than required 
for clear extraction of asymmetries of the size indicated by the HERMES
results, the trends in the COMPASS data are clearly consistent with the
the HERMES data discussed above. With continued running very much improved
precision is expected for the COMPASS measurements.

\subsection{Related measurements}

Because the single spin asymmetries measured in SIDIS are determined by
the product $\delta q\cdot H^{\perp}_{1}$, an independent determination
of $H^{\perp}_{1}$ will greatly facilitate the extraction of $\delta q$.
Measurements of spin-dependent fragmentation functions
are planned by the BELLE experiment at the KEKB asymmetric
collider by studying azimuthal angle correlations in $e^{+}e^{-}
\rightarrow \pi^{+}\pi^{-}X$ reactions. The analysis in progress is
focused on the separation of Collins asymmetries from physics background.

At RHIC measurements of the Sivers and Collins effects are planned in
$p\uparrow p$ reactions. Significant single spin asymmetries 
already have been reported by the
STAR collaboration for forward-produced $\pi^{0}$'s at $\sqrt{s}$=200 GeV
and $x_{q}\ge$0.6. Star will measure $A_{N}^{\pi^0}$ in a different
kinematic region, midrapidity where $x_{q}\approx 0.1$.

\section{Discussion of Results}

Perhaps the most interesting feature of the 
HERMES data is the unexpectedly large
negative $\pi^-$ Collins azimuthal moment. As a result of the proof
of factorization\cite{ji} for spin-dependent DIS cross sections that involve
in leading twist the effects of intrinsic $p_T$ or fragmentation $K_T$, 
the single-spin asymmetries can be expressed in terms of the usual product
of distribution ({\it e.g.} transversity) and fragmentation 
({\it e.g.}Collins) functions. Thus, a reasonable leading order 
ansatz for the the Collins asymmetry is  
\begin{equation}
A^{h}_{UT}\propto \,S_{\perp}\frac{\sum_{q,\overline{q}}e^{2}_{q}\delta 
q(x)H_{1}^{\perp q}(z)}{\sum_{q',\overline{q'}}e^{2}_{q'} 
q'(x)D_{q'}^{h}(z)}
\label{exeq3}
\end{equation}
where $e_q$ is the electric charge of the quark of flavor $q$.
The negative $\pi^-$ moment would be explained by a substantial disfavored
Collins function that 
describes the fragmentation of $u$ quarks to $\pi^-$ mesons 
with the opposite sign to that of the favored function. Because the 
Collins function acts as an analyzing power for transverse quark 
polarization, either sign for its amplitude is possible. 

A simple model the Collins fragmentation process\cite{artu} based on the Lund
semiclassical colored string mechanism provides a plausible explanation
for the opposite signs of the favored and unfavored fragmentation functions.
\begin{figure}[h]
\begin{center}
\includegraphics[width=15cm]{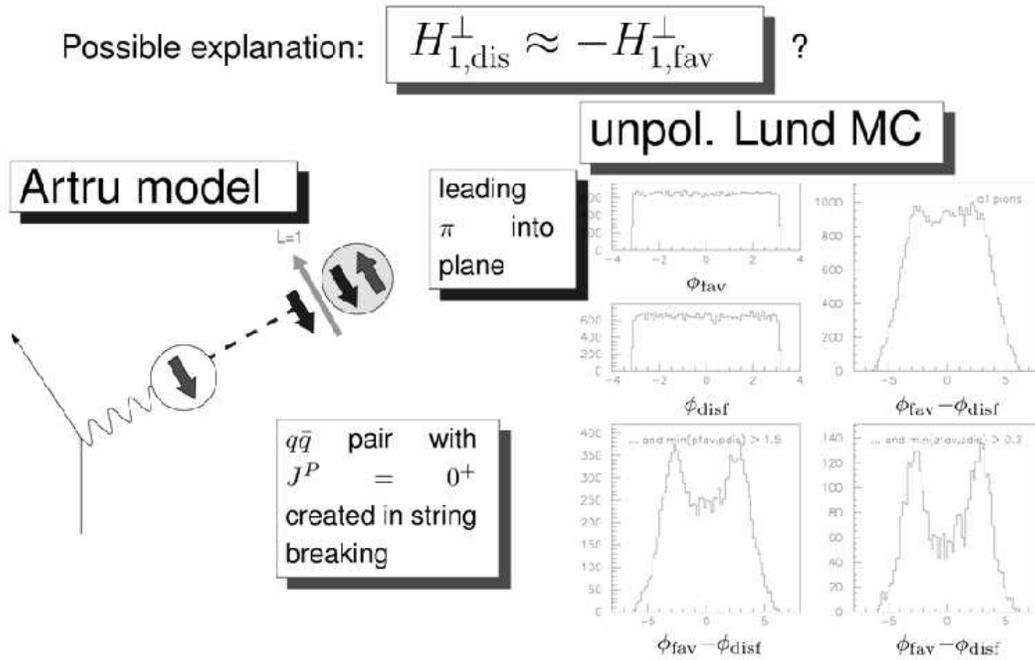}
\end{center}
\caption{\label{string} String model of spin dependent fragmentation of
X. Artu {\it et al.} \cite{artu}. The sections at the right present the 
results of a Lund Monte Carlo simulation which shows the back to back
emission of fragmentation pions at high $z$.
 }
\end{figure}
The implications of this model for pion electroproduction are illustrated
in Figure \ref{string}. With the absorption of the virtual photon, the spin 
of the struck quark is flipped and the string connecting the quark to the
remanent diquark is stretched until the first string break as illustrated in
the Figure. It is assumed in the model that the $q\overline{q}$ pair 
formed at the break is produced in a triplet P state, {\it e.g.} $J^P = 0^+$.
As indicated in this configuration the leading pion is preferentially produced
with momentum into the plane. The quark spins at the break will have spin
orientations opposite to those of the emerging struck quark. Applying the 
same model at the second string break where a disfavored pion is formed, 
the disfavored pion will be produced with transverse momentum opposite in
direction to that of the leading pion. This anticorrelation in $P_{\pi\perp}$
between the favored and unfavored pions is demonstrated by the 
{\sc Jetset} simulation shown at the right of Figure \ref{string}. The results
of this model which is based on Lund string fragmentation show, particularly
for $z_{favored}>0.2$,  a strong back to back azimuthial correlation between
favored and unfavored pions. 

In theoretical representation of the single spin azimuthal asymmetry as 
measured, a $p_T$-dependent density function ({\it e.g.} Sivers) or a
$k_T$-dependent fragmentation function ({\it e.g.} Collins) appears
inside a convolution integral over $p_T$ and $k_T$. When asymmetry data
is averaged over $|P_{h\perp}|$, the deconvolution of distribution and 
fragmentation functions can only be made with some assumption about their
dependence on $p_T$ and $k_T$, respectively. However it has been 
demonstrated \cite{kotzinian} that a model independent extraction can
be made by weighting the experimental cross sections by $|P_{h\perp}/z|$
in the integral over this quantity, prior to any moment analysis.
With this weighting the Collins asymmetry takes the form
\begin{eqnarray}\label{eqn:conv1} \nonumber
\bigg\langle \frac{|\mb{P}_{h\perp}|}{(z M_h)} \sin{(\phi{+}\phi_S)}
\bigg\rangle_{UT}\!\!\!\!\! (\xbj,y,z)
 &\equiv \frac{\int  \de \phi_S \de^2\! \mb{P}_{h\perp}
\,{|\mb{P}_{h\perp}|/(z M_h)}\,\sin{(\phi{+}\phi_S)}\; 
\de^6\sigma_{UT}}
{\int  \de \phi_S \de^2\! \mb{P}_{h\perp}\, \de^6\sigma_{UU}}  \\
&= |\mb{S}_{\st}|\,\frac{
         B(y)\,\sum_q e_q^2\, {\delta q (\xbj)}\,
           {H_1^{\perp (1)q}(z)}}
{A(y)\,\sum_q' e_q'^2\, {f_1^{q'} (\xbj)}\,
         {D_{q'}^{h}(z)}} ,
\end{eqnarray}
where including the $1/z$ in the weight relates the asymmetry to the first
$z$-moment of the Collins function. For the Sivers asymmetry the form is
\begin{eqnarray}\label{eqn:conv2} 
\bigg\langle \frac{|\mb{P}_{h\perp}|}{(z M_p)} \sin{(\phi{-}\phi_S)}
\bigg\rangle_{UT}\!\!\!\!\! (\xbj,y,z)
= |\mb{S}_{\st}|\,\frac{
         \sum_q e_q^2\, {f_{1T}^{\perp q} (\xbj)}\,
           {D_{q}^{h}(z)}}
{\sum_{q'} e_{q'}^2\, {f_1^{q'} (\xbj)}\,
         {D_{q'}^{h}(z)}}
\end{eqnarray}
where including the $1/z$ in the weight relates the asymmetry to the first
$x$-moment of the Sivers function. 
$M_h$ is the hadron mass, $M_p$ is the proton
mass, and the
superscript(1) indicates the $p_T$- or $k_T$-moment of the distribution or
fragmentation function, respectively.  
These equations form the basis for subsequent analysis to extract 
$\delta q(x)$ from the Collins moment and
${f_{1T}^{\perp q} (\xbj)}$ from the the Sivers moment.   

Extraction of the flavor-dependent Sivers function is more accessible than
that of transversity, because the required unpolarized fragmentation functions 
$D_{q}^{h}(z)$ are known independently. A purity technique, previously used
to extract longitudinal helicity distributions \cite{airapet} can be 
applied. From equation \ref{eqn:conv2} the Sivers asymmetry takes the form
 \begin{eqnarray} \label{eqn:purity1} 
  A^{h}_{S}(x)  = -|S_{T}| \frac{\int_{z_{min}}^{1}dz\sum_{q}e^{2}_{q}
  {q(x)}\cdot{zD^{h}_{q}(z)}}
  {\int_{z_{min}}^{1}dz\sum_{q'}e^{2}_{q'}
  {q'(x)}\cdot{D^{h}_{q'}(z)}}
  \cdot{\frac{\delta q(x)}{q(x)}} 
    =  \sum_{q}{P^{h}_{q}(x)}
  \frac{\delta q(x)}{q(x)}
  \end{eqnarray}
where the hadron quark purity $P_{q}^{h}(x)$ is the probability that 
a quark $q$ was struck in an event $e^{+}\, +\,N\,\rightarrow e^{+'}  
+ h + X$. It is a spin-independent quantity which can be calculated
from a Monte Carlo simulation of the fragmentation process. One can rewrite
equation \ref{eqn:purity1} in a matrix form as
\begin{equation} {\vec{A}=\left( \begin{array}{c}
                       A^{h_1}(x) \\ ... \\ A^{h_n}(x)
                   \end{array} \right)},
     {\vec{Q}=\left( \begin{array}{c}
                       \delta q_{1}(x)/q_{1}(x) \\ ... \\
                       \delta q_{1}(x)/q_{n}(x)
                    \end{array} \right)},
     {P=[P^{h}_{q}(x)]},\,\, {\vec{A}(x)}={P}{\vec{Q}(x)}
\end{equation}
where $A(x)$ becomes a vector whose elements are all the integrated 
measured asymmetries which are to be included in the analysis. The 
$\vec{Q}(x)$ vector contains the quark transversities. These quantities
are now connected by the purity matrix which contains effective integrated
purities. The determination of the quark transversities from the 
experimentally determined measured azimuthal asymmetries is reduced to
the task of inversion of equation \ref{eqn:purity1} to obtain $\vec{Q}(x)$.
Such an analysis of the HERMES data is in progress. In principle, a 
similar procedure can be used to extract transversity from the Collins
asymmetries, however, in this case the calculation of the purities is
prevented at present by the lack of reliable estimates of the Collins
fragmentation functions.  

\begin{figure}[h]
\begin{center}
\includegraphics[width=14cm]{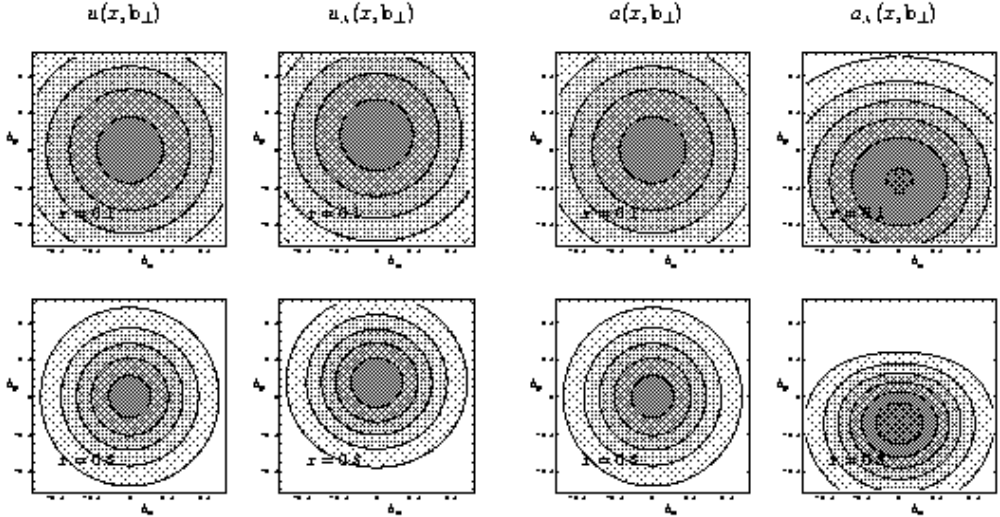}
\end{center}
\caption{\label{impact} Parton distributions in impact parameter space
from ref. \cite{burkhardt}. For each plot the grayscales are normalized
to the central value. For each quark flavor the left column is for a
longitudinally polarized nucleon and the right column for a nucleon
polarized in the $x$ direction.
 }
\end{figure}

An interesting conjecture based on the use of generalized parton distributions
to determine the spacial distribution of partons in the transverse
plane \cite{burkhardt} indicates that Sivers asymmetries for $u$ and $d$ 
quarks should have opposite signs. In this model the scattering process
is treated in terms of an impact parameter ($b_X$,$b_y$) formalism. The
left-right asymmetry in the hadron emission is attributed to final
state interactions of the struck quark as it leaves the target. These
interactions with a parton distribution which has been distorted by
quark orbital motion leads to the asymmetry. Examples of the distortion as
represented in the distribution of parton density in impact parameter
space is shown in Figure \ref{impact}.

\section{Summary and outlook}

The first measurements of asymmetries directly related to transversity
and the Sivers function have been made. From the data on the Collins
asymmetry the flavor-disfavored Collins fragmentation function appears to
be opposite in sign to the favored one and large. The nonzero Sivers
asymmetries can be interpreted as a manifestation of quark orbital 
angular momentum. The results to date provide a demonstration of a new
tool for studying the transverse structure of the nucleon. Such measurements
will provide estimates of the nucleon's transversity distribution
and promise to shed light on the relativistic properties of the 
confined states of quarks and gluons.

\begin{center}{\bf Acknowledgements}\end{center}

It is a pleasure to acknowledge the work of the HERMES transversity group
and, in particular, that of Ulrike Elschenbriosh, Ralf Seidl, Gunar
Schnell, and Naomi Makins who furnished much of the source material for this
paper. This work is supported in part by funds provided by the US
Department of Energy, Office of Nuclear Physics, under contract 
W-31-109-ENG-38.

\end{document}